\documentclass[aps,prb,preprint,superscriptaddress,amsmath,floatfix,showpacs]{revtex4}
\usepackage{graphicx}  
\usepackage{subfigure}

\begin{document}
\def\bar{\begin{eqnarray}}
\def\ear{\end{eqnarray}}
\def\beq{\begin{equation}}
\def\eeq{\end{equation}}
\newcommand{\degrees}{\ensuremath{^{\circ}}}
\title{Plasmonic Resonances and Electromagnetic Forces Between Coupled Silver Nanowires }
\author{Klaus Halterman}\email{klaus.halterman@navy.mil}
\affiliation{Physics and Computational Sciences, Research and Engineering Sciences Department, Naval Air Warfare Center,
China Lake, California 93555}
\author{J. Merle Elson}
\affiliation{Physics and Computational Sciences, Research and Engineering Sciences Department, Naval Air Warfare Center,
China Lake, California 93555}
\author{Surendra Singh}
\affiliation{Department of Electrical Engineering, University of Tulsa, Tulsa, Oklahoma 74104}
\date{\today}



\begin{abstract}
We compute the electromagnetic response and corresponding forces between two silver nanowires. 
The wires are illuminated by a plane wave which has the electric field vector perpendicular to the axis of the wires,
insuring that plasmonic resonances can be excited. We consider a nontrivial square cross section geometry 
that has dimensions on the order of $0.1 \lambda$, where $\lambda$ is
the wavelength of the incident electromagnetic field.
We find that due to the plasmonic resonance, there occurs great
enhancement of the direct and mutual electromagnetic forces that are exerted on the nanowires.
The Lippman-Schwinger volume integral equation 
is implemented to obtain solutions to Maxwell's equations for various $\lambda$ and 
separation distances between wires. The forces are computed using Maxwell's stress tensor and
numerical results are shown for both on and off resonant conditions. 
\end{abstract}
\maketitle

\section{Introduction}
The phenomenon of surface plasma resonance excitation (surface plasmon) was first predicted \cite{ritchie} 
as an explanation of observed energy loss spectra of electron beams penetrating thin metal films. Also, at 
optical wavelengths the reflectance of metal (Al, Ag, Au) mirrors with surface roughness is significantly 
affected by surface plasmon excitation \cite{stanford}.  When considering scattering of plane waves at wavelength 
$\lambda$ by particles of dimension $<< \lambda$, Rayleigh scattering theory can often be applied. In this 
case the shape of the particle is not important and the scattering cross section is typically quite small. 
However more recently, the study of electromagnetic enhancement and field localization by nano-sized metallic 
structures via plasmonic resonance has been the focus of research activity. 
In this case, when  plasmonic excitation is involved, the shape and material details of the structure become important with respect to conditions for resonance and this greatly increases the scattering of electromagnetic waves.  When two or more nanosized structures are illuminated at a resonant wavelength, the electromagnetic coupling and associated charge redistribution that occurs can result in a significant mutual coupling force. Plasmon resonances lead to extremely large localized fields at specific wavelengths in the the vicinity of nanoparticles, thereby resulting in a large scattering cross section \cite{r02a} and mutual forces between nanoparticles. The localized fields also play an important role in the surface enhanced Raman scattering. This enhancement can be large enough to enable the detection of a single molecule. A primary driving force for the recent focus on plasmon resonance of nanoparticles is the application of optical properties of these structures in novel optical devices\cite{r02,r03} and biosensors \cite{r04,r05}. It has been noted that nanowires with different cross sectional shapes such as, cylindrical, triangular and elliptical exhibit multiple resonances depending on their particular shape. For example, the field enhancement 
in the vicinity of a 20 nm triangular particle can exceed 400 times the incident field amplitude, while it is only 10 for a cylindrical particle of the same size \cite{r06}. In the past few years, several researchers have investigated the resonance effects of particle sizes larger than 100 nm. Very recently, attention has been focused on resonance effects of nanoparticles in the 20 to 50 nm range \cite{r06,f1}.

A primary difficulty in dealing with particle size smaller than 100 nm, especially during resonance conditions, 
is the increasing computational complexity due to rapid variations of the field over short distances.
For structures that deviate from the simpler cylinder and sphere geometries,
analytical approaches are not available, and one must rely on an effective numerical approach that correctly incorporates the necessary boundary conditions. An efficient numerical procedure for the computation of fields at the plasmon resonance wavelength of 
nanoparticles of arbitrary cross section is afforded by the Lippman-Schwinger integral equation \cite{r02a}. 
This numerical formalism is a volume integral equation utilizing the Green's tensor for 2D or 3D geometries. 
In this paper, we calculate the electromagnetic forces exerted on two nanowires with square cross section. 
The side dimensions of the nanowires are 30 nm and the wavelength of the incident field varies from 300-516 nm.  

\section{method}
We first outline the volume integral equations, which are appropriate for the 
investigation of nanosystems, and in this case electromagnetic resonances in 2D coupled nanowire 
structures. The system is assumed infinite in the $z$ direction, and
all spatial variation occurs in the $x$-$y$ plane. The physically relevant quantities that we shall focus  
on are the electric ($\bf E$) and magnetic ($\bf H$) fields and the nature of net forces exerted on a given wire 
under  conditions of plasmonic resonance. 
In the absence of magnetic media ($\mu=1$), 
the basic integral equations are written as \cite{blackbook},
\begin{subequations}
\begin{align}
\label{LS.1}
{\bf E}({\bf r}) &= {\bf E}^{\rm inc}({\bf r})+k_0^2  \int_S d^2r^{\prime} {\bf G}_e({\bf r},{\bf r}^{\prime}) \cdot \delta\epsilon({\bf r}^{\prime}){\bf E}({\bf r}^{\prime}), \\
\label{LS.2}
{\bf H}({\bf r}) &= {\bf H}^{\rm inc}({\bf r})-i k_0 \int_S d^2r^{\prime} {\bf G}_m({\bf r},{\bf r}^{\prime}) \cdot \delta\epsilon({\bf r}^{\prime}){\bf E}({\bf r}^{\prime}),
\end{align}
\end{subequations}
where $k_0=\omega/c$, ${\bf r}=(x,y)$, and the integration is over the cross-sections $S$ of the 
nanowires, which are embedded in vacuum.  The free-space electric   
dyadic Green's function, ${\bf G}_e({\bf r},{\bf r}^\prime)$, is a second-rank tensor and 
its explicit form has been given elsewhere \cite{blackbook}.  The
magnetic dyadic Green's function, ${\bf G}_m({\bf r},{\bf r}^{\prime})$, is calculated using 
${\bf G}_m({\bf r},{\bf r}^{\prime})=\nabla\times\bigl[{\bf I}\, g_0 ({\bf r},{\bf r'})\bigr]$,
where $g_0 ({\bf r},{\bf r'})=(i/4) H_0(k_0 \rho) \exp(i k_z z)$.
Since the incident field propagates solely in the $x-y$ plane ($k_z=0$), this yields 
the following nonzero components:
$G^m_{xz}=-i/(4\rho) k_0 (y-y') H_1(k_0\rho)$ and
$G^m_{yz}=i/(4\rho) k_0 (x-x') H_1(k_0\rho)$,
where $\rho = |{\bf r} - {\bf r}^{\prime}|$.
The other components are  found straightforwardly using the antisymmetric relation $G^m_{i j}=-G^m_{j i}$.
The scalar permittivity contrast is mapped according to $\delta\epsilon({\bf r}) \equiv \epsilon - 1$, when ${\bf r}$ is 
entirely within the nanowire with permittivity 
$\epsilon$, and $0$ otherwise.  We then cast the integrand of (\ref{LS.1}) into the form of a 
linear matrix equation system by discretizing the wire surface as a fine grid for numerical integration. Once the electric field is  found within the scattering medium, ${\bf E}({\bf r})$ is then inserted into (\ref{LS.2}) to calculate ${\bf H}({\bf r})$. As a check, we calculated the ${\bf H}$ field directly from the ${\bf E}$ field using the Maxwell-Faraday law and found consistency with the results obtained using Eq. (\ref{LS.2}).

To calculate the net force acting on a particular wire, we begin with
Maxwell's stress energy tensor, $\bf T$, which has components,\cite{jacko}
\beq
T_{\alpha \beta} = \frac{1}{4\pi}\bigl[E_\alpha D_\beta + H_\alpha B_\beta-\frac{1}{2}
({\bf E \cdot D}+{\bf B \cdot H} )\delta_{\alpha \beta} \bigr].
\eeq
The net force on an object, enclosed by an area $A$ is then given by 
${\bf F} = \langle\oint_A ( {\bf T \cdot n} )\, dA \rangle$
where ${\bf n}$ is the outward normal to the surface $A$ enclosing a given nanowire,
and $\langle \rangle$ denotes the time average. For
the geometry under consideration, and with ${\bf E}=(E_x,E_y)$,  only
the $T_{xx}, T_{xy}$, $T_{yx}$ and $T_{yy}$ components of the stress tensor contribute to $\bf F$.

\section{numerical results}

The  Lippmann-Schwinger equation (Eq. (\ref{LS.1})) can be cast into the standard form 
of a linear equation system $\bf A x=b$, and is solved here numerically by  
an efficient and stabilized version of the conventional BiConjugate  
gradient method.\cite{bcg} The matrix ${\bf A}$ does not have to be  stored in computer memory, allowing discretization of  the scatterer  over a very fine scale without undue limitations  on the storage requirements.  We model the permittivity of the metal nanowires using the experimental silver data in Ref.~\onlinecite{palik}.  The square nanowires are taken to be $30$ nm per side dimension and the grid count per nanowire is $80 \times 80$. This yields a grid spatial resolution of $0.375 \,{\rm nm} \times 0.375 \,{\rm nm}$. 
Several numerical tests confirmed that this resolution is quite satisfactory. 

In general for silver nanowires with small cross sections, the inverse collision time for electrons can increase from the additional scattering events that take place at the surface, and thus scattering from the boundary becomes relevant. This problem has been treated using alternate techniques. The effect of electron screening is expected to also reduce the inverse collision time, however a careful modeling of the particle surface showed \cite{apell} an effective scattering rate  that is comparable to the classical result.

Significant attraction between two metal nanowires can be generated by Coulombic
forces. If the gap distance is such that the plasmonic evanescent
near-fields overlap, then the potential for coupled-wire resonant modes
exist.  These coupled-wire modes can be very complex depending on gap
distance, permittivity, wavelength,  and geometric parameters. These modes
can have a widely varying spatial distribution of the electron plasma. It
then follows that if the gap fields are enhanced because of a
resonance condition, there should be enhanced Coulombic forces.

\begin{figure}
\centering
\includegraphics[width=4in]{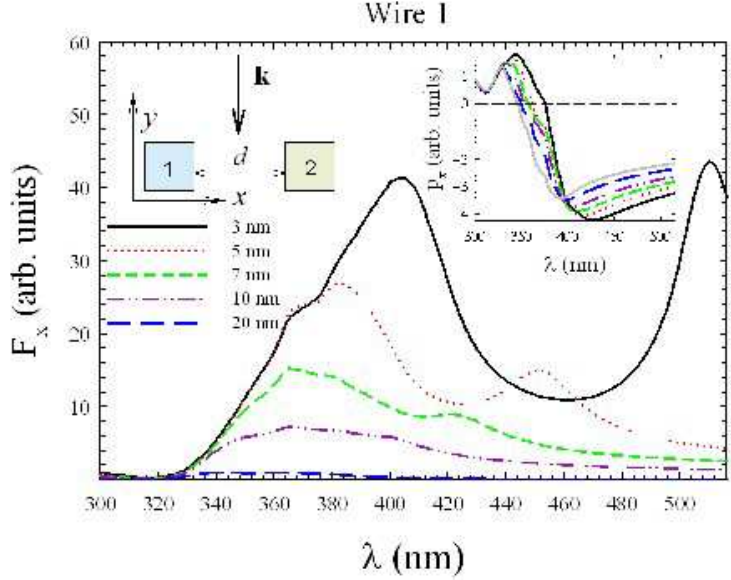}
\caption{(Color online) The wavelength dependence of the $x$-component of the time-averaged total force on the left wire (wire 1) 
for five different separations $d$, as indicated in the legend.
The incident plane wave propagates downwards in the $-\hat{\bf y}$ direction, while the field 
is polarized along the $\hat{\bf x}$ direction, which is appropriate to excite surface plasmons. 
The force is always positive,
indicating attraction between the nanowires since the adjacent wire (wire 2) feels 
an identical force that is opposite in sign (not shown).
The inset illustrates the corresponding $x$-component of the 
dipole moment, with the solid gray curve representing the case of a single wire.
}
\label{fig1} 
\end{figure}

To address this, we begin by examining the forces exerted on the nanowires due directly to the 
incident wave and the mutual coupling between the wires.  The results 
naturally depend on the propagation direction of the incident field, given by its wavevector 
${\bf k} = -k_0(\cos\theta \hat{\bf x} + \sin\theta \hat{\bf y})$. In Fig.~\ref{fig1}, 
we illustrate the $x$-component of the time-averaged force, $F_x$,  on the left wire (wire 1)  
as a  function of the incident field wavelength, $\lambda$, and for various gap
distances $d$.  The  incident electric field vector is polarized along the $\hat{\bf x}$ 
direction, and  the wavevector is directed along $-\hat{\bf y}$. This particular 
incident field propagation direction yields oppositely directed net forces on each wire that 
are equal in magnitude.  It is evident  that $F_x$ is positive for wire 1 over the
whole $\lambda$ range shown, indicating attraction among the nanowires. 
It is seen that for much of the $\lambda$ range, smaller gaps yield
greater maximum values of $F_x$. All curves show a decrease and blueshift
in the $F_x$ peak as $d$ increases, except for the  $d = 20$ nm curve, which
shows very little apparent interaction between the wires. This is
consistent with a coupled-wire resonance condition that is supported by
the $p_x$ inset  showing the $x$-component of the dipole moment of wire 1. 
This quantity is 
obtained by integrating the appropriate component of the real part of the polarization,  
${\bf P}({\bf r}) \equiv 1/4\pi \delta\epsilon({\bf r}) {\bf E}({\bf r})$, over the cross 
sectional area of the wire. 
We show here only $p_x$ since this
component of the dipole moment encapsulates the relevant charge distribution involved in interwire coupling.
In
the inset the solid gray curve shows for reference, $p_x$ for a
single isolated wire.  Note that the $d = 20$ nm curve for $p_x$ is
very similar to the solid gray curve, further indicating very
little mutual interaction between the wires.  
For $330 \,\,{\rm nm}\lesssim \lambda \lesssim 400$ nm,
$p_x$ has negative slope and becomes more negative quite rapidly with
wavelength.  
This is approximately the region over which the $F_x$ curves reach their primary peak.
It is also interesting that for larger wavelengths, there is another set
of  $F_x$ peaks that are visible in the $d =
3$, $5$, and $7$ nm curves. These peaks become washed out at larger $d$, and 
approximately correlate
with $p_x$ (inset) where the
slopes are positive.

\begin{figure}
\centering
\subfigure
{
    \label{vec:sub:a}
    \includegraphics[width=4.4in]{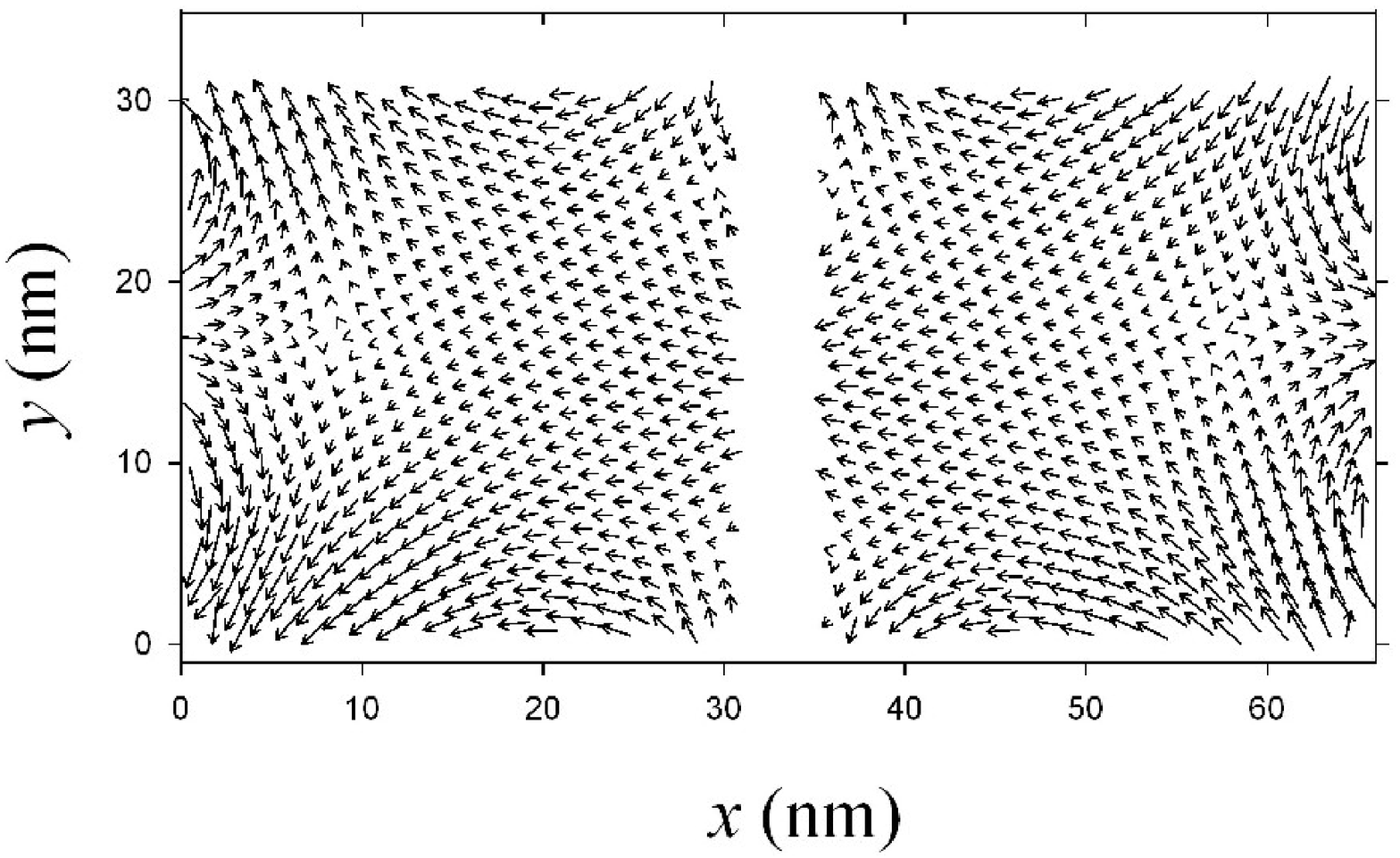}
}
\\ \vspace{-.3cm}
\subfigure
{
    \label{vec:sub:b}
    \includegraphics[width=4.4in]{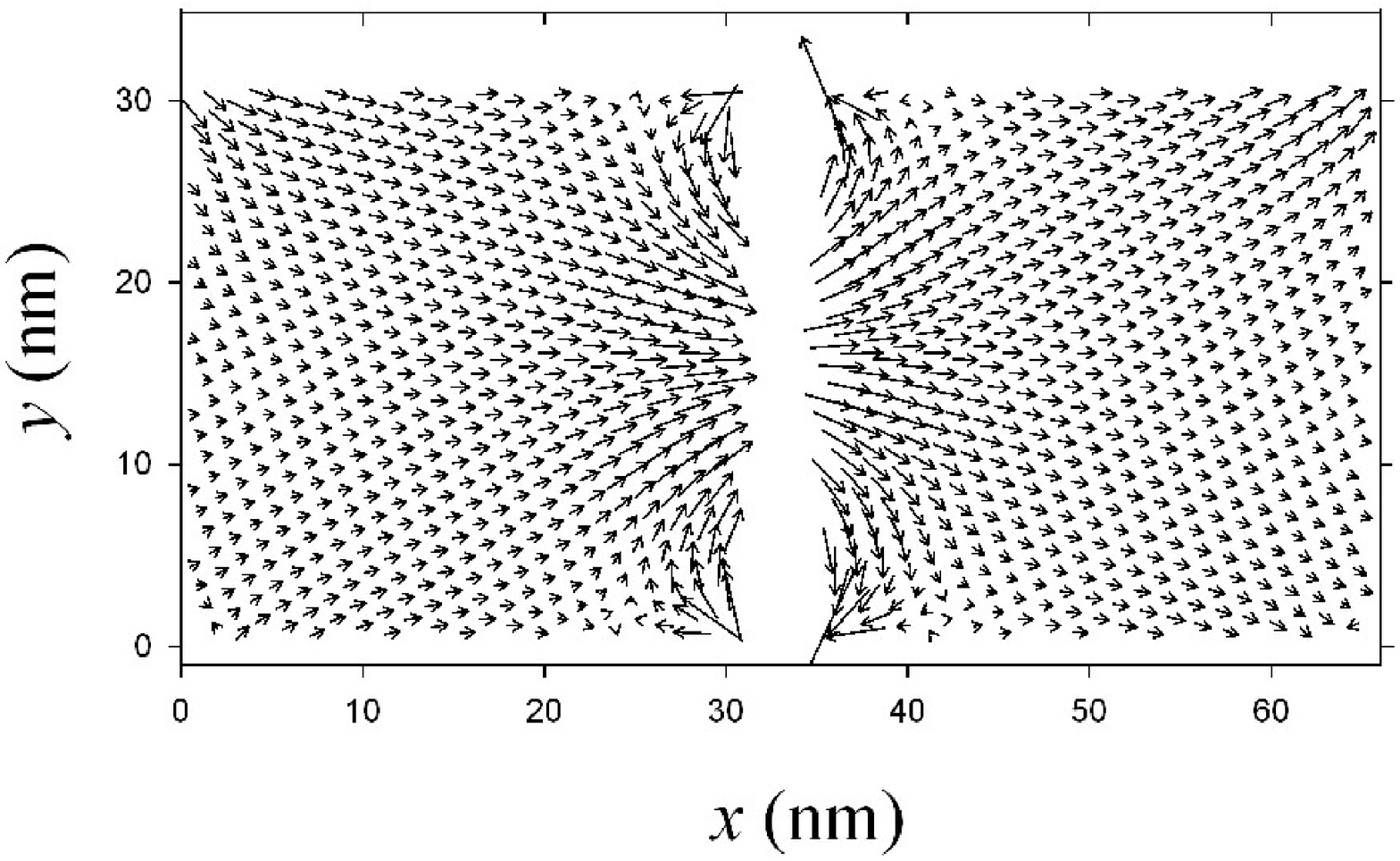}
}

\caption{Vector plots of the $\bf E$ field within the nanowires 
illustrating the local field directions. Here an incident
plane wave is traveling downwards as in Fig.~\ref{fig1}. The wires
are separated by $5$ nm. The top panel corresponds to  $\lambda=340$ nm
while the bottom panel is for $\lambda=385$ nm. 
The top panel clearly shows that the $\bf E$ field is most intense at the outer corners opposite
the gap. Integration of the polarization over the left wire yields $p_x>0$ as shown
in Fig.~\ref{fig1}. Conversely, in the bottom panel the
field in proximity to the gap is most intense and integration
over wire 1 yields $p_x<0$, again consistent with Fig.~\ref{fig1}.
This demonstrates the reason for the relatively large and attractive 
forces between the wires at
$\lambda=385$ nm compared to $\lambda=340$ nm.
}
\label{vec:sub} 
\end{figure}

The sign change in the dipole moment exhibited in Fig.~\ref{fig1} is easily visualized by means of 
vector plots. To this end we show in Fig.~\ref{vec:sub}
the electric field vector distribution within the nanowires. 
The wires are separated by $5$ nm and the top and bottom panels correspond to 
incident field wavelengths of $\lambda=340$ nm and $\lambda=385$ nm respectively.
In the top $\lambda=340$ nm panel, the corner dipole fields are most intense 
on the outer left and right sides.
There is a subsequent nonuniform, strongly position dependent field orientation within the nanowires.
The $\bf E$ field on the sides nearest the gap region is relatively small compared to the corner fields. 
In the bottom $\lambda=385$ nm panel,
the situation is markedly different. Here the gap field is dominant
and different mode patterns arise. These results are consistent with Fig.~\ref{fig1},
where $F_x$ at $\lambda=340$ nm is small compared to $\lambda=385$ nm.

\begin{figure}
\centering
\vspace{-.3in}
\subfigure
{
    \label{fig2:sub:a}
    \includegraphics[width=3.5in]{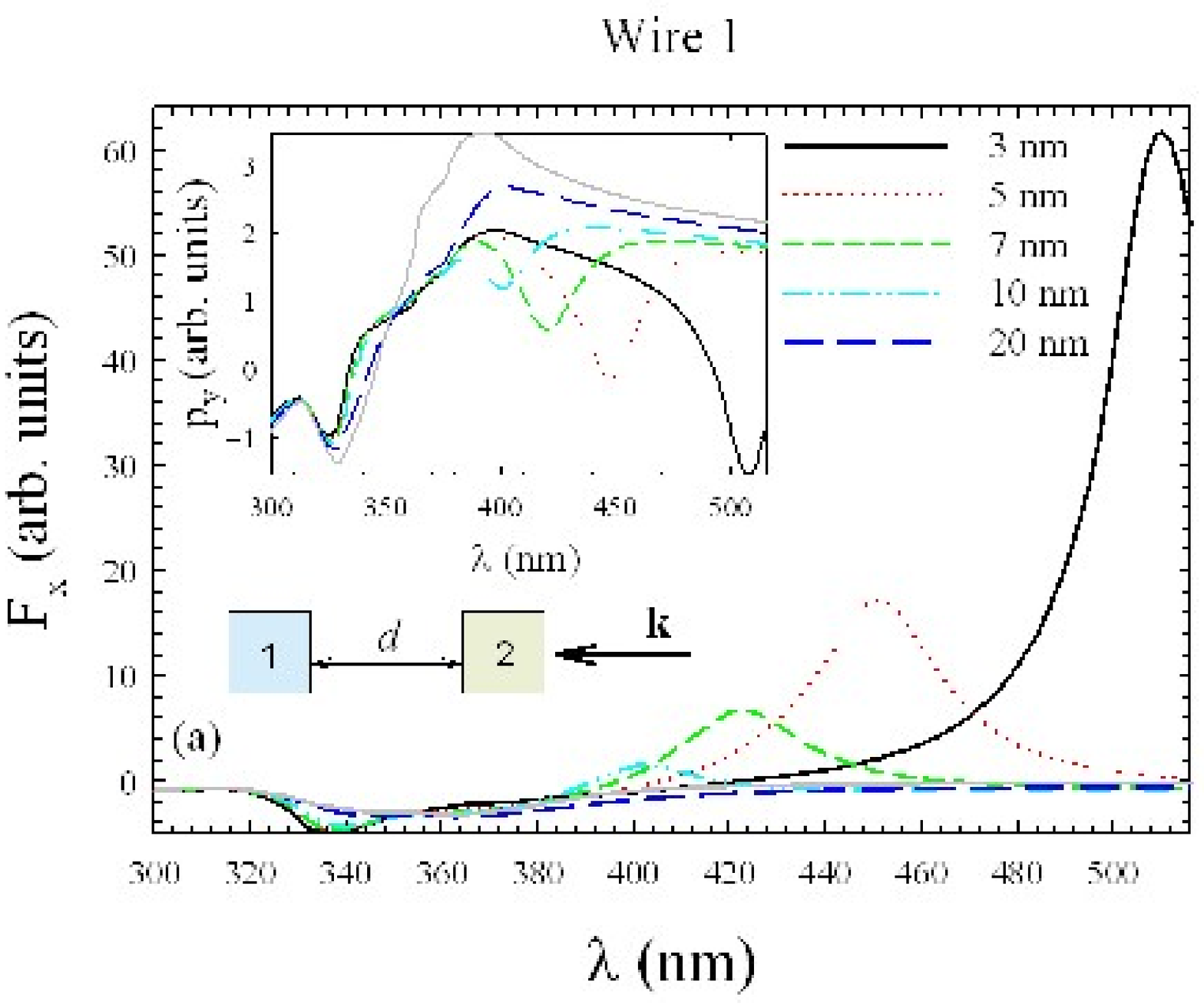}
}
\\ \vspace{-.2cm}
\subfigure
{
    \label{fig2:sub:b}
    \includegraphics[width=3.5in]{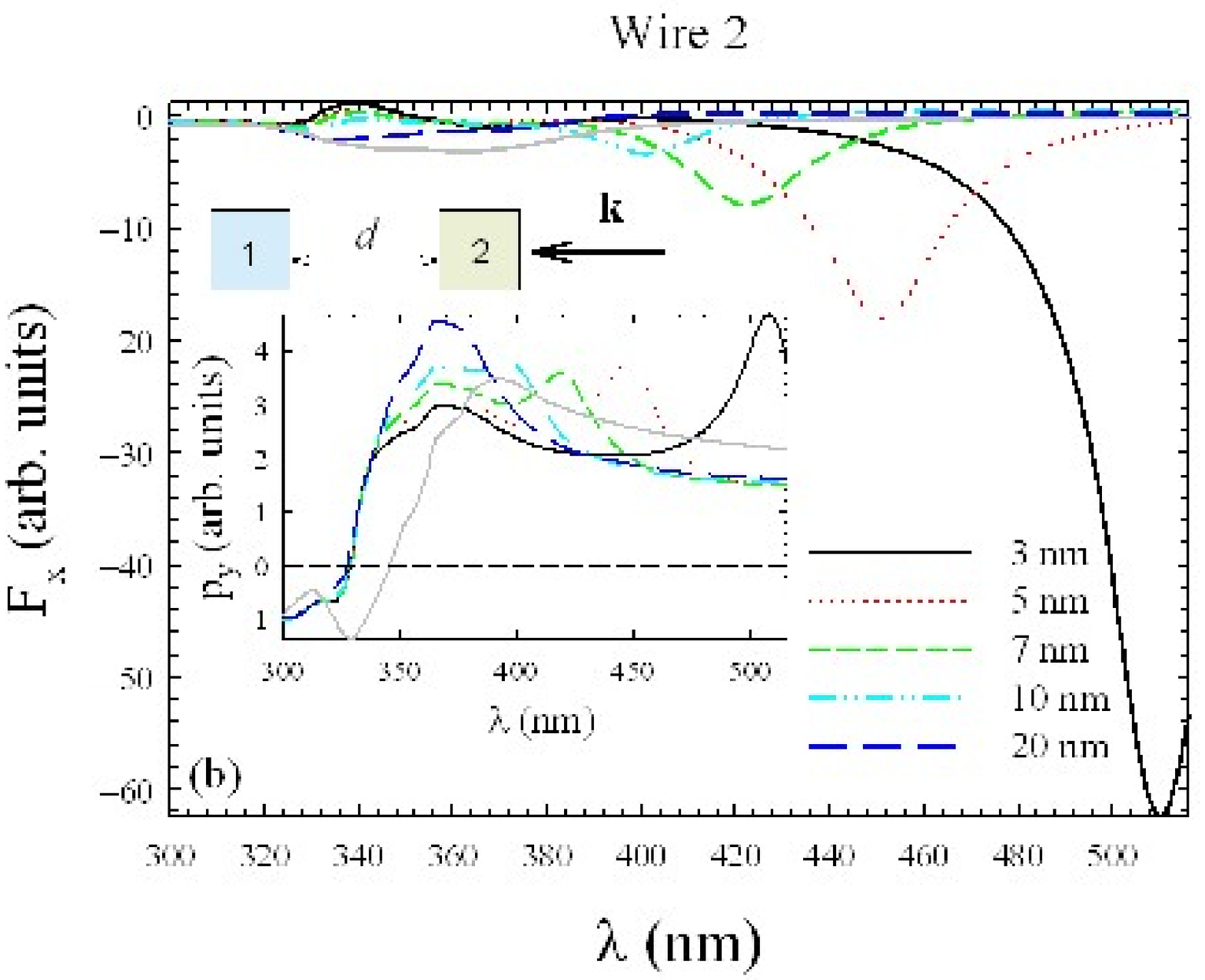}
}

\caption{(Color online) The wavelength dependence of the $x$-component of the time-averaged total force on  wire 1 (a) and wire 2 (b) 
for five different separations $d$, as indicated. 
The coordinate system is identical to that shown
in Fig.~\ref{fig1}. The incident plane wave propagates with ${\bf k}$ 
in the $-\hat{\bf x}$ direction and the electric field is polarized in the $\hat{\bf y}$ direction. 
The dashed curves correspond to $F_x$ on a single nanowire.
There are also 
insets that show the net dipole moment $p_y$ for each nanowire. 
For this incident field orientation $p_x$ is zero.
For reference, the solid gray curves in the
two insets shows $p_y$ for an individual nanowire. Note that in the neighborhood of $\lambda \approx 336$ nm, $p_y<0$ 
in (a) and $p_y>0$ in (b) which yields a repulsive force between the nanowires. When $\lambda \gtrsim 400$ nm, the 
wires are largely attractive, except for $d=20$ nm, where there is large enough separation that the 
interaction between wires is minimal.
}
\label{fig2:sub} 
\end{figure}

We now vary the incident electric field so that it is now polarized in the $\hat{\bf y}$ direction, and 
propagates along $-\hat{\bf x}$. We show in Fig.~\ref{fig2:sub:a}, $F_x$
on  wire 1 as a function of $\lambda$. Figure ~\ref{fig2:sub:b} depicts the same
quantities for wire 2. In general, the net force on each wire is no longer equal in magnitude  
due to the uneven radiation effects directly from the incident wave. 
Figure \ref{fig2:sub:a} shows that
for sufficiently small wire separations ($d \leq 10$ nm), the 
force on wire 1 changes sign in a continuous manner as $\lambda$ is varied. 
The force on wire 2 is mainly negative except for a narrow band of $\lambda$
in the vicinity of $\lambda\approx 336$ nm (see Figure \ref{fig2:sub:b}),
which correlates to when the force on wire 1 is most negative.
Thus the wires experience a {\it repulsion} at these wavelengths and separations.
For larger separations
$F_x$ diminishes and tapers towards zero at higher wavelengths, indicating that electromagnetic 
coupling between the two wires is negligible. 
The plasmon resonances within these structures reveal themselves through
prominent peaks in the force curves in both figures. 
The peak positions are strongly dependent on $d$, as they broaden and 
reduce in magnitude as $d$ increases.
Comparing the scales in Figs.~\ref{fig1} and \ref{fig2:sub}, we find that the 
overall resonance effects observed for the forces at small separations is reduced somewhat for this incident 
field direction.
The insets in Fig.~\ref{fig2:sub} show  
the $y$-component of the electric dipole moment ($p_y$) versus $\lambda$ for wires 1 and 2.
For wire 1 and  $d=5$ nm, Fig.~\ref{fig2:sub:a} shows that $p_y$ undergoes a minimum at $\lambda\approx328$ nm and
then sharply crosses zero at $\lambda\approx 332$ nm. This trend is
similar for the other separations, with only a slight redshifting for larger $d$. 
For both wire 1 and wire 2 a clear correlation exists between $p_y$ and the associated force; 
for small enough $d$, the pronounced peaks or depressions in $p_y$ correspond to when
$|F_x|$ is largest. These features therefore become redshifted and spread-out with decreasing separation. 
As one would expect for $d=20$ nm, the general features of $p_y$ approach that
of the limiting case of a single nanowire.

\begin{figure}
\centering
\vspace{-1.1in}
\includegraphics[width=5in]{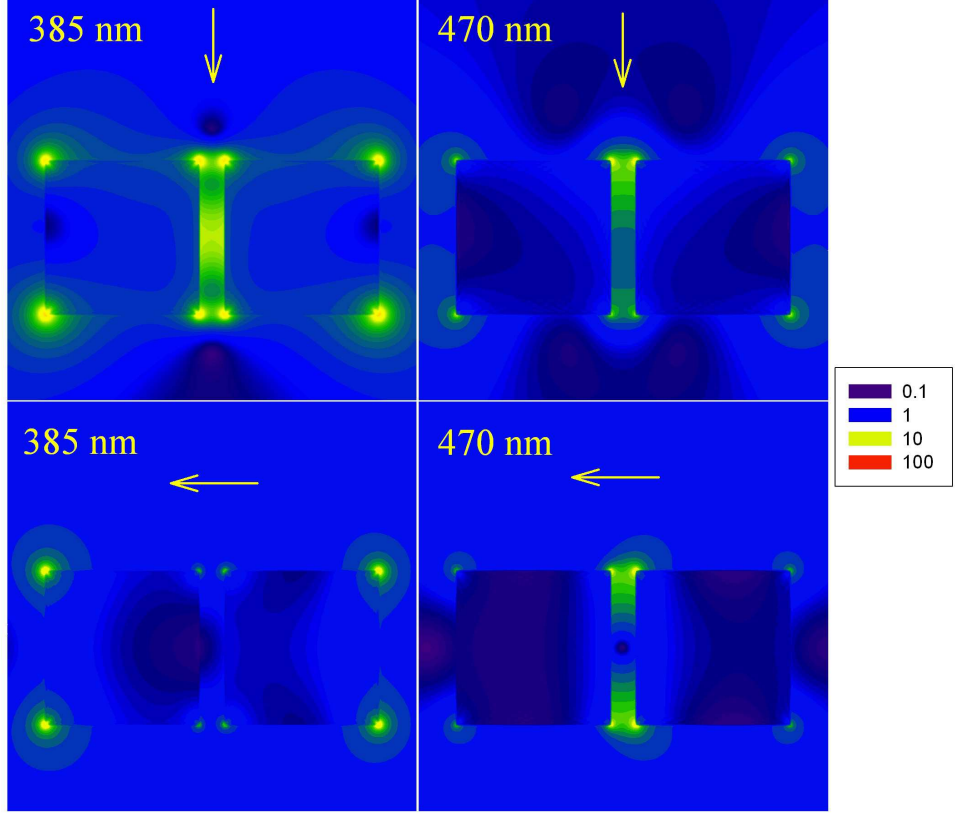}
\caption{(Color online) Electric field amplitude contours (normalized by the incident field) for two nanowires 
illuminated by a plane wave with wavevector direction 
indicated by 
the arrows. The separation distance is $d=5$ nm. The electromagnetic response of the nanowire pair is quite different 
depending on the angle of incidence and wavelength. In the two upper plots, the electromagnetic response and subsequent attractive force 
is greatest for the shortest wavelength. Conversely, in the lower plots the response is greatest for the longer 
wavelength.  These results are  consistent with Figs. \ref{fig1} and \ref{fig2:sub}. }
\label{onefield} 
\end{figure}

To contrast the near-field electric field patterns
for differing incident wave directions, we display in Fig.~\ref{onefield} the
electric field amplitudes in the region of the two nanowires.
The wires are 
set at a fixed distance of 5 nm and are illuminated at two different
wavelengths.
The top two panels correspond to an incident field
directed downwards (as shown by the arrows) with $\lambda=385$ nm and $\lambda=470$ nm.
The bottom two panels are for an incident electric field traveling towards the left.
Beginning with the top left panel, at $\lambda=385$ nm there is a large field enhancement 
from the corners and within the gap between the wires. This is consistent with
Fig.~\ref{fig1}, where the force has a maximum at that wavelength. As the wavelength is increased
to $\lambda=470$ nm, the system is no longer at resonance and the interaction between the wires
has decreased. This follows from  Fig.~\ref{fig1}, where the force is reduced by
nearly a factor of three from its peak value.
A different situation is observed in the bottom panels, where at $\lambda=385$ nm
the electric field is small between the wires, although the outer corners are still illuminated.
At $\lambda=470$ nm, the field amplitude is intensified predominantly within the gap as the nanowires 
are electrically coupled. There is a clear relationship between the force profiles in Fig.~\ref{fig2:sub}
and the field amplitude plots. It is apparent that the incident field 
``sees" a different geometry
based on its orientation relative to the coupled structure; downward propagating waves
interact with an effectively larger nanowire in addition to the interwire gap that may support additional
modes due to the wire-wire interaction.

\begin{figure}
\centering
\subfigure
{
    \label{fx365:sub:a}
    \includegraphics[width=3.5in]{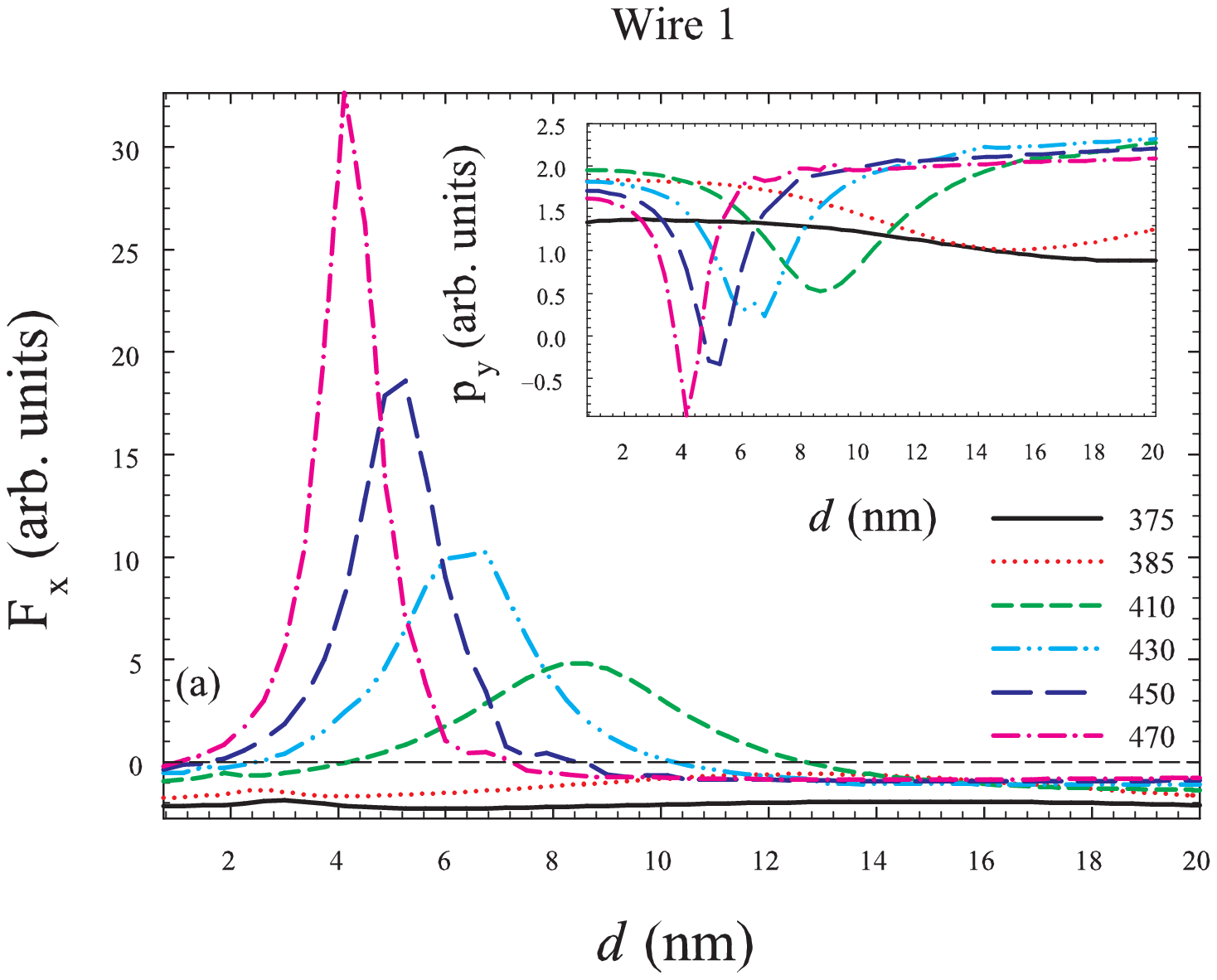}
}
\\ \vspace{-.2cm}
\subfigure
{
    \label{fx516:sub:b}
    \includegraphics[width=3.5in]{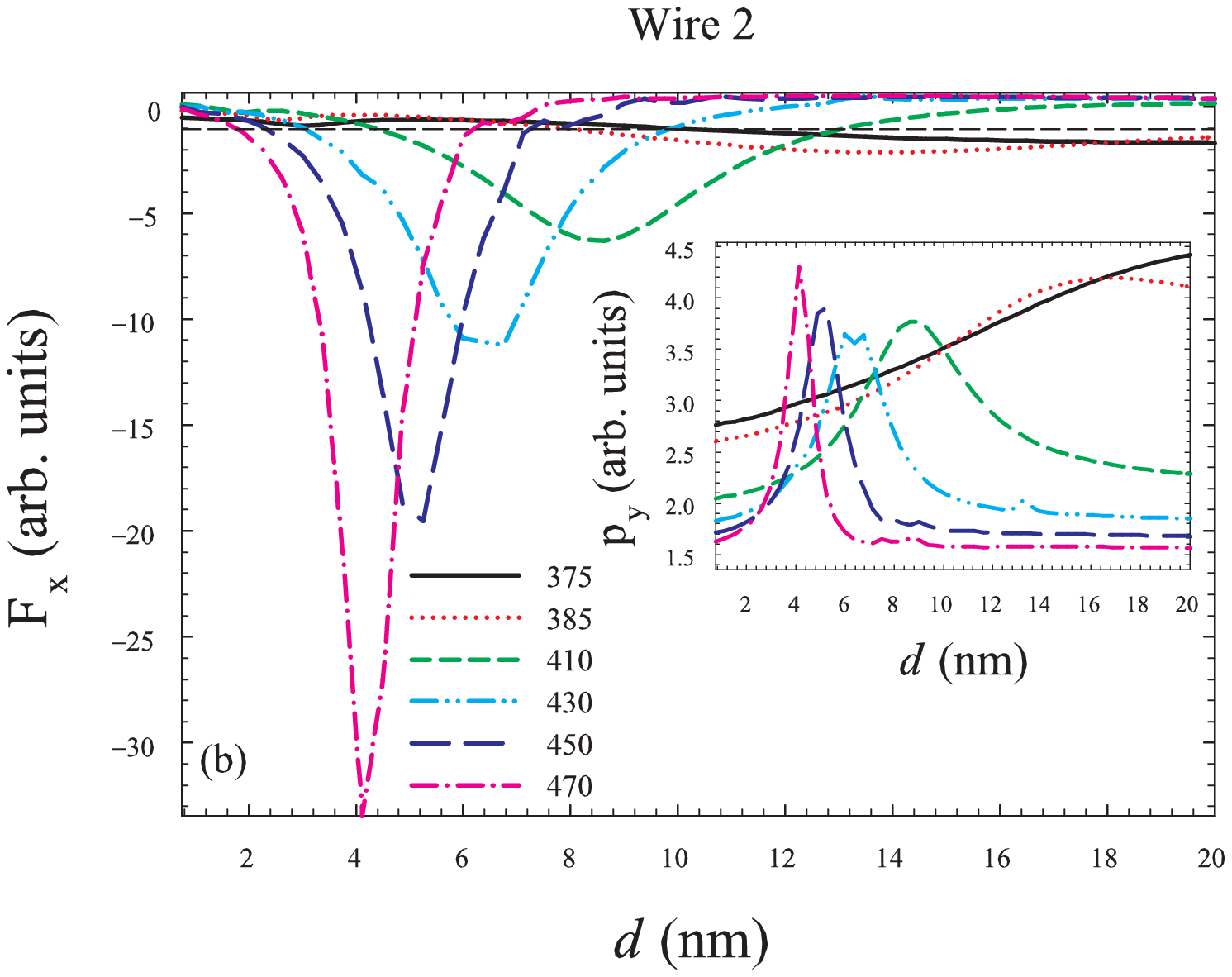}
}

\caption{(Color online) The $x$-component of the total force $F_x$ as a function of wire separation $d$ on wire 1 (a) and wire 2 (b) 
for a field incident 
from the right ($-\hat{\bf x}$ direction) as in Fig.~\ref{fig2:sub}.
Six different wavelengths are shown, and labeled in the legends. While the wires experience no net force in the $y$-direction, 
the forces in the $x$-direction vary greatly with wavelength and separation $d$. It is seen that depending 
on the wavelength, there are resonant peaks in the attractive forces between the two wires in the $\lambda$ range
$410-470$ nm. For the shorter wavelengths, $\lambda = 375$ and $385$ nm, there is 
relatively little response of the wire pair when compared to the longer wavelengths. The inset 
plots show the net induced $y$-component of the dipole moment, $p_y$, corresponding to each force curve.}
\label{fxboth:sub} 
\end{figure}

Next, in Fig.~\ref{fxboth:sub}, the force on each of the two  nanowires is shown as a function of the separation 
distance $d$, and for
an incident field directed along $-\hat{\bf x}$. Also shown is the corresponding induced net dipole moment, $p_y$, 
for each wire. There are six $\lambda$ considered and the variation of $F_x$ with $d$ is strongly 
wavelength dependent. Consider first the longest wavelength, $\lambda=470$ nm where the separation range 
$d = 4 - 5$ nm  shows a large positive $F_x$ in  Fig.~\ref{fxboth:sub}(a) and a correspondingly large 
negative $F_x$ in  Fig.~\ref{fxboth:sub}(b).  This indicates a resonant condition resulting in a 
relatively strong attractive force. This is in contrast to the wavelengths, $\lambda=375$ and 385 nm where 
the $F_x$ curves remain negative and comparatively featureless over the entire $d$ range. Clearly no resonant 
condition occurs for this shorter wavelength as is also evidenced
in Fig.~\ref{fig2:sub}. These results are consistent with the two lower plots 
in Fig. \ref{onefield} that show little gap field enhancement at $\lambda=385$ nm and a large 
gap field for $\lambda=470$ nm. The inset  plots in Fig.~\ref{fxboth:sub} are similar in that the 
maxima in $|F_x|$ show up as discernible peaks in $|p_y|$,
reflecting the  charge redistribution. 
In general there is a redshift in the peak values and a 
sharp decline of $|F_x|$ as $d$ increases, indicating a corresponding shift in the  
resonance parameters of the mutually interacting wires.

\begin{figure}
\centering
\includegraphics[width=4in]{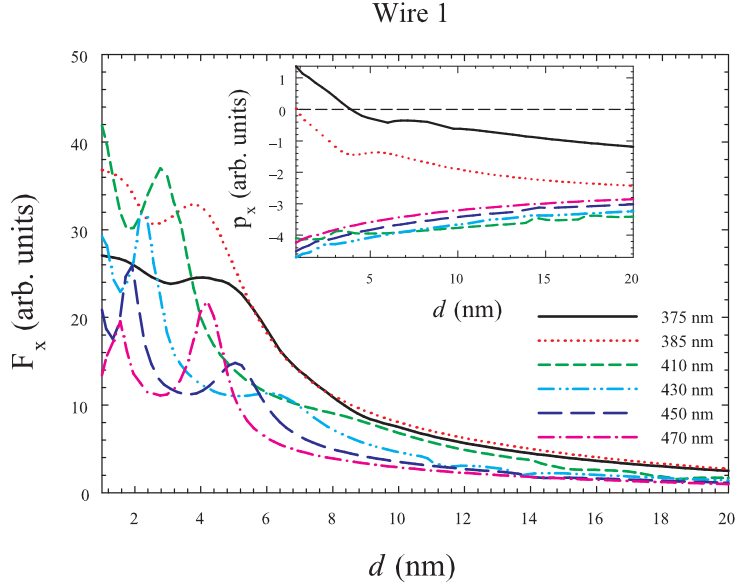}
\caption{The normalized $x$-component of the force
on wire 1 versus separation $d$.
The force on the adjacent wire (wire 2) is identical except
for a sign change.
The incident
field is directed downward as seen in the Fig.~\ref{fig1}. 
The inset is the corresponding (normalized) dipole moment and the curves
correlate with the legend and
the same set of
parameters are used as in the main plot.
}
\label{fig4} 
\end{figure}

The normalized force, $F_x$, on 
the left wire (wire 1) as a function of
$d$ for an incident field 
propagating downwards (in the $-\hat{\bf y}$ direction) 
is shown in Figure \ref{fig4}. This figure
is analogous to Fig.~\ref{fxboth:sub}.
As in Fig.~\ref{fig1}, the force on each wire is equal in magnitude and oppositely directed,
yielding an attractive nature over the relatively broad wavelength range studied. 
The main resonant peaks are present at small separations and then 
$F_x$ monotonically declines
towards zero with increased $d$. 
The wavelengths that give the largest $F_x$ occur in the
range $385-410$ nm. Outside of that range, increasing
$\lambda$ has the effect of reducing  the force and
narrowing its peaks. 
For sufficient separation distances, the force 
becomes weakly dependent on the wavelength. 
These characteristics reveal that there
are optimal separations and wavelengths to utilize the plasmon driven
force enhancement.  
The inset shows the normalized $x$-component of the dipole moment which
is identical in each nanowire.
The curves each follow a similar trend, with
large separations maintaining a negative dipole moment, while 
for $\lambda=375$ nm the possibility exists to
induce a positive net dipole moment
for small separations. Thus for a particular value of $d$ it is possible to have
a vanishing dipole moment in the wire. This follows from 
regions within the wire structure that have  oppositely oriented
polarization vectors that sum to zero.

In conclusion, we have studied the electromagnetic response of two long silver nanowires  
illuminated by a plane wave with the electric field perpendicular to the axis of the wires. The
volume integral approach used here provides accurate 
solutions to Maxwell's equations over length scales that are
much smaller than the wavelength of the driving field.
We found that
by varying the wavelength, separation 
distance, and angle of incidence of the electric field, 
the collective oscillation of the electrons induced
appreciable forces and coupling in the nanowires.
The results shown here can play a role in
the fabrication and design aspects
of optical spectroscopy and waveguiding 
over nanometer length scales.

\acknowledgements
This work of K. H. and J.M.E. is funded in part by the Office of Naval Research 
(ONR) In-House Laboratory Independent Research (ILIR) Program and by a grant of HPC resources from 
the Arctic Region Supercomputing Center at the University of Alaska Fairbanks 
as part of the Department of Defense High Performance Computing Modernization 
Program. S.S. would like to acknowledge support from the ONR/ASEE Summer Faculty Program.


\begin{thebibliography}{99} 

\bibitem{ritchie} R. H. Ritchie, Phys. Rev. {\bf 106}, 874 (1957).
\bibitem{stanford} J. L. Stanford, J. Opt. Soc. Am. {\bf 60}, 49 (1970).
\bibitem{r02a} J. P. Kottmann, O. J. F. Martin, D. R. Smith, and S. Schultz, Phys. Rev. B {\bf 64} 235402 (2001).
\bibitem{r02} M. Quinten, A. Leitner, J. R. Krenn and F. R. Aussenegg, Opt. Lett. {\bf 23} 1331 (1998).
\bibitem{r03}  J. C. Weeber, A. Dereux, C. Girard, J. R. Krenn and J. P. Goudonnet, Phys. Rev. B {\bf 60} 9061 (1999).
\bibitem{r04} C. Viets and W. Hill, J. Raman Spectroscopy {\bf 31} 625  (2000).
\bibitem{r05} R. Elghanian, J. J. Storhoff, R. C. Mucic, R. L. Letsinger and C. A. Mirkin, Science {\bf 277} 1078 (1997).
\bibitem{r06} J. P. Kottman and O. J. F. Martin, Opt. Express {\bf 8}, 655 (2001).
\bibitem{f1} J. R. Arias-Gonzalez and M. Nieto-Vesperinas, J. Opt. Soc. Am. A {\bf 20}, 1201 (2003).
\bibitem{blackbook} J. J. H. Wang, {\it Generalized Moment Methods in Electrodynamics} (Wiley, Toronto, 1991).
\bibitem{jacko} J.D. Jackson, {\it Classical Electrodynamics}, 2nd ed. (Wiley, New York, 1975).
\bibitem{bcg} Fletcher, R.,{\it Conjugate Gradient Methods for Indefinite Linear Systems, Lecture Notes in Mathematics 506}, (Springer-Verlag Berlin, 1976) pgs. 73-89.
\bibitem{palik} E. D. Palik, {\it Handbook of Optical Constants of Solids} (Academic Press, Washington, D.C., 1985).
\bibitem{apell} P. Apell and D. R. Penn, \prl {\bf 50} 1316 (1983).

\end{thebibliography}
\end{document}